\documentclass[12pt]{iopart}
\usepackage[utf8]{inputenc}
\usepackage{graphicx}
\usepackage{iopams}  
\usepackage{verbatim}
\usepackage[colorlinks=true,linkcolor=blue]{hyperref}

\graphicspath{{./figures/}}
\begin{document}

\title{Complex phase space of a simple synchronization model}

\author{Sz. Horv\'at and Z. N\'eda}

\address{Babe\c{s}--Bolyai University, Dept.~of~Theoretical and Computational Physics \\
str. Kog\u{a}lniceanu, nr. 1 \\
400084 Cluj-Napoca, Romania
}
\ead{szhorvat@phys.ubbcluj.ro}

\begin{abstract}
The phase-space of a simple synchronization model is thoroughly investigated. The model considers two-mode stochastic oscillators, coupled through a pulse-like interaction controlled by simple optimization rules. A complex phase space is uncovered as a function of two relevant model parameters that are related to the optimization threshold and the periods of the two oscillation modes. Several phases with different periodic global output signals are identified. It is shown that the system exhibits partial synchronization under unexpectedly general conditions. 
\end{abstract}

\submitto{\NJP}

\pagestyle{plain}
\tableofcontents

\maketitle


\newcommand{\tmax}{\ensuremath{t_{\mathrm{max}}}}


\section{Introduction}

Spontaneous synchronization appears in a large variety of systems in nature.  Well-known examples include biological systems such as fireflies flashing in unison or crickets chirping together \cite{Sismondo1990}, rhythmic applause \cite{Neda2000,Neda2000a}, pacemaker cells in the heart \cite{Peskin1975}, the menstrual cycles of women \cite{Stern1998}, oscillating chemical reactions, mechanically coupled metronomes, pendulum clocks hung on the same wall, and many other systems.

Several mathematical models have been proposed to explain and describe the spontaneous synchronization phenomena in large interacting ensembles.  Most of these models can be grouped in one of two broad categories that are distinguished by the nature of coupling between the oscillators: those that are based on phase coupling and those that are based on a pulse-like coupling.

The prototypical model for phase coupled oscillators is the Kuramoto model \cite{Kuramoto1987}.  The Kuramoto model consists of an ensemble of globally coupled rotators where the state of each unit is described by a $\theta \in [0, 2\pi)$ periodic phase variable.  When not coupled to the others, a single oscillator rotates with its natural angular frequency, $\omega$.  The angular frequencies of the oscillators are distributed according to a unimodal distribution $g(\omega)$.  In the presence of coupling the equation of motion of an oscillator is 
\[
\frac{d \theta_i}{dt} = \omega_i + K \sum_{j=1}^{N} \sin (\theta_j - \theta_i), 
\quad i = 1, \cdots, N,
\]
where $K$ is a coupling constant.  This interaction naturally leads to synchronization because it tries to minimize the phase difference between oscillators.  The synchronization level of oscillators can be characterized by the order parameter $r \in [0,1]$ defined by the equation $r e^{i \varphi} = \frac{1}{N} \sum_{j=1}^N e^{i \theta_j}$.  

The specially chosen trigonometric form of the coupling makes it possible to study this model using analytic methods.  Kuramoto showed that in the limit of a very large number of oscillators, there exists a critical value of the coupling constant, $K_c$, so that if $K < K_c$ then the phases of oscillators are distributed randomly ($r=0$), 
while if $K > K_c$, then the oscillators become partially synchronized ($r > 0$) \cite{Kuramoto1987}.

Many systems in nature however can't be assigned an associated periodic phase variable, thus the Kuramoto model is not a suitable description for them.  In the case of systems where the interaction between oscillators is pulse-driven (such as in the case of fireflies, firing of neurons, rhythmic clapping), \emph{integrate and fire} type synchronization models are used \cite{FitzHugh1961,Nagumo1962,Bottani1996,Pikovsky1997}.  The integrate and fire model is based on the assumption that each oscillator has a monotonically increasing state variable.  When the state variable reaches a threshold value, the oscillator ``fires'': it emits a pulse, and its state variable is re-set to zero.  When an oscillator detects a pulse in the system, its state variable suddenly increases by a constant value $K$, which may cause it to reach its own threshold.  It is easy to see that in this system the firing of an oscillator may trigger an avalanche of pulses, causing many oscillators to fire within a short period of time.  As a result of this, synchronization will emerge. The order parameter used to characterize the synchronization level of these systems is usually defined as the size of the largest avalanche compared to the total number of units present. Similarly to the case of the Kuramoto model, it can be shown that in the case of global coupling there is a critical value of the constant $K$ above which the oscillators will fire in a partially synchronized manner.

Both of these basic model categories have many variations, where other interactions are also considered or in which the coupling is not global.  Their statistical behaviour and  the appearance of synchronization depends on the strength as well as the topology of the coupling.  There are a number of review works available on the topic of the statistical mechanics of spontaneous synchronization \cite{Strogatz2000,Strogatz2003,Pikovsky2001}.

A new, simple model that leads to synchronization in a non-trivial manner was recently introduced by Nikitin et al. \cite{Nikitin2001,Neda2003}.  It was inspired by the study of rhythmic applause \cite{Neda2000,Neda2000a}.  Modifications of this basic model were studied in several subsequent papers \cite{Neda2003,Sumi2009,Horvat2009}, by using numerical modelling or an experimental realization of the system \cite{Sumi2009}.
In this model, the oscillators are coupled through emitted pulses, similarly to integrate and fire type models. The interaction between the oscillators does not however favour synchronization in a direct way.  The considered coupling intends to keep the average output in the system close to a threshold level.  Unexpectedly, synchronization appears as a side effect of this optimization interaction.  In this model the oscillators are bimodal, being able to operate in either a slow or fast (long or short period) oscillation mode, and therefore contribute to the total output of the ensemble with a higher or lower average output.  At the beginning of each oscillation period, an oscillator unit will decide whether to choose the slow or fast mode depending on the detected total output level in the system.  The periods of the oscillation modes are also randomly fluctuating.

Previous numerical studies performed on different variations of this model have shown that for certain parameter values the ensemble of bimodal oscillators will partially synchronize and produce a periodic output signal \cite{Nikitin2001,Neda2003,Sumi2009,Horvat2009}.  The studies were performed as a function of the randomness of the periods of the modes and the threshold value. The effect of changing the ratio of the periods of the two modes was however not investigated in detail, and the phase space of the model was not mapped with high accuracy before.  In the present work we focus on exploring the behaviour of the model as a function of the threshold level and the ratio of mode periods, and explore the phase space with a much higher accuracy than before.  We have found that this simple model of bimodal oscillators has a phase space with a complex and surprisingly non-trivial structure.

\section{The two-mode stochastic oscillator model}

\subsection{Description of the model}

The basic version of the model considers an ensemble of $N$ identical bimodal, globally coupled, stochastic oscillators \cite{Nikitin2001}.  At any time, an oscillator can either be active, emitting a signal of strength $1/N$, or inactive, emitting no signal.  Therefore the total output level of the system can vary between 0 and 1.  These oscillators can be intuitively thought of as flashing units.  For simplicity, from now on we shall refer to active ones as \emph{lit} and inactive ones as being in an \emph{unlit} or \emph{dark} state.  In accordance with this intuitive picture, the sum of the units' output levels can be thought of as the total light intensity in the system.

The units are stochastic bimodal oscillators. They can operate in two oscillation modes, one with a shorter and one with a longer period.  These will be referred to as mode 1 and mode 2, respectively.  The periods of the modes are random, and their mean values are denoted by $\tau_1$ and $\tau_2$.

An oscillation period consists of three phases, $A$, $B$ and $C$.  During phase $A$ and $B$ the units are dark, while during phase $C$ they are lit.  The duration of phase $A$, $\tau_A$, is a random variable drawn from the interval $[0, 2\tau^*]$ with a uniform distribution.  The mean value of $\tau_A$ is $\langle \tau_A \rangle = \tau^*$. In this paper we shall assume that $\tau^* \ll \tau_1$.  The duration of phase $B$, $\tau_B$, can have two values, $\tau_{B1}$ and $\tau_{B2}$, corresponding to the two oscillation modes.  The duration of the lit phase, $\tau_C$, is fixed.  The average lengths of the periods of the modes is the sum of the mean durations of these three phases: $\tau_1 = \langle \tau_A \rangle + \langle \tau_{B1} \rangle + \langle \tau_C \rangle = \tau^* + \tau_{B1} + \tau_C$ and similarly $\tau_2 = \tau^* + \tau_{B2} + \tau_C$. Since the units stay lit for a greater fraction of the short period mode than the long one, the average light intensity will be larger when the units are oscillating in the short period mode.

The coupling between the oscillators is realized through an interaction that strives to optimize the total light intensity in the system, denoted $f$.  At the beginning of each period, a unit decides which mode to follow based on whether the total light intensity, $f$, is greater or smaller than a threshold level $f^*$:
\begin{itemize}
  \item If $f \le f^*$, the shorter period mode will be chosen.  Since an oscillating unit stays lit for a greater fraction of a full period when it is operating in the short mode, this will help in increasing the average light intensity in the system.

  \item If $f > f^*$, the longer period mode will be chosen, reducing the average total light intensity in the system.
\end{itemize}

By this dynamic, each oscillating unit individually aims to achieve a total output intensity as close to $f^*$ as possible, based on their instantaneous measurements of the output level.  As a side effect of this optimization procedure, synchronization can emerge: the total output intensity of the system becomes a periodic function and the units will flash in unison \cite{Nikitin2001,Neda2003,Sumi2009,Horvat2009}.

The simple model presented in the previous paragraphs differs from the original one described in \cite{Nikitin2001} and \cite{Neda2003} only in the distribution of the duration of the stochastic phase, $\tau_A$.  In the original model, $\tau_A$ was exponentially distributed, and the behaviour of the system was studied as a function of the variables $\tau^* = \langle \tau_A \rangle$ and $f^*$.  The present paper focuses on the case when $\tau^* \ll \tau_1$, therefore the precise statistical distribution of $\tau_A$ does not influence the results significantly.  The reason for choosing a uniform distribution for this study is that using a distribution defined on a bounded interval simplifies numerical modelling of the system.  In this paper the system is studied as a function of the parameter $f^*$ and the ratio of the average periods of the two oscillation modes, $\tau_2 / \tau_1$.  

There are several variations possible on the basic version of the model. Some of these variations have been previously shown to also lead to synchronization.  In \cite{Horvat2009} a version of the model with a duration dark phase and a variable duration lit phase was studied, while in \cite{Sumi2009} it was shown that synchronization emerges also when using multimodal oscillators.  The lit phase can occur at the beginning or at the end of the oscillation period, leading to different behaviours.  Finally, in this paper we will show that synchronization will occur even when a reversed optimization is used, that aims to achieve an output as different from the threshold $f^*$ as possible.

Three versions of the bimodal oscillator model will be considered: {\bf model 1.}\ the basic model described above with a fixed-duration lit phase, and a variable duration dark phase;  {\bf model 2.}\ a model with variable duration lit phase and a fixed-duration dark phase; and {\bf model 3.}\ fixed-duration lit phase and variable duration dark phase with a reversed choice of the long or short modes depending on the $f^*$ value.  This last case will be referred to as ``anti-optimization'' because the oscillators strive to achieve an output as different from $f^*$ as possible.  Partial synchronization will emerge in all three cases.

\subsection{The order parameter}

We need a quantitative measure to characterize the synchronization level of the system.  The order parameter used in previous studies \cite{Nikitin2001,Neda2003,Sumi2009} measures the periodicity level of the output signal.  Let the output signal be denoted by $f(t)$, and define the function
\[
\Delta(T) = \frac{1}{2M} \lim_{x \rightarrow \infty} \frac{1}{x},
  \int_0^x | f(t) - f(t + T) | \, dt
\]
where:
\[
M = \lim_{x \rightarrow \infty} \frac{1}{x} 
\int_0^x | f(t) - \langle f(t) \rangle | \, dt
\]
Here $\langle f(t) \rangle$ denotes the mean value of the function $f(t)$ over the interval $[0, \infty]$.   Normalization by $2M$ ensures that the value of $\Delta(T)$ will be between 0 and 1.  It is clear that if $f(t)$ is a perfectly periodic signal, then $\Delta(T) = 0$ for all integer multiples of the period, as well as $T = 0$.  If the signal is approximately periodic, then the function $\Delta(T)$ will have minima at integer multiples of the period.  The location of the first deep minimum, $T_m$, will correspond to the period of the signal.  The level of periodicity can then be characterized by $p = 1/\Delta(T_m)$.  Note that the value of $p$ can vary between $1$ and $\infty$.

When using numerical modelling to simulate the system, the output level is computed at discrete points in time.  Unfortunately the periodicity measure $p$ turned out not to be practical when the output signal is highly periodic and is known at discrete points only.  The finite time resolution limits the precision of finding $T_m$, which in turn might have a significant effect on the computed value of the periodicity level $p$.  The behaviour of $p$ as a function of the model parameters will no longer be characteristic of the  dynamics, but will reflect the discretization of time variables.  This becomes apparent only when modelling a larger number of oscillators than has been done previously and sampling the parameter space with a higher resolution.

Therefore, here we chose a different order parameter to characterize the synchronization level in the system. It has been observed in previous numerical studies that when synchronization emerges, the amplitude of the total light intensity function, $f(t)$, will be high as well.  Therefore it is possible to use the amplitude of the signal to detect partial synchronization.  It is practical to use the standard deviation of the signal to characterize its amplitude.  The standard deviation is defined as follows:
\[
\sigma = \lim_{x \rightarrow \infty} \sqrt{ \frac{1}{x} 
\int_0^x (f(x) - \langle f(x) \rangle)^2 \, dx }
\]
where
\[
\langle f(x) \rangle = \lim_{x \rightarrow \infty} \frac{1}{x} \int_0^x f(x) \, dx.
\]
This proved to be a robust measure that is not sensitive to outliers in the signal and characterizes intuitively well the ``flashing'' behaviour of the system.  A nonzero $\sigma$ value corresponds to a partially synchronized flashing dynamics.

\section{Details of the computer simulation method}

The oscillator system was studied by extensive computer simulations. The value of the $\sigma$ order parameter was investigated as a function of the threshold parameter, $f^*$, and the ratio of the average periods of the two oscillation modes, $\tau_2/\tau_1$.  The period of the output signal was estimated using a simple auto-correlation method, and the order parameter, $\sigma$, was calculated from a time average on an interval given by a large \emph{integer} number of periods (when the signal was periodic), for increased accuracy.

\subsection{Efficient simulation methods}

Previous studies  \cite{Nikitin2001,Neda2003,Sumi2009,Horvat2009} used a direct method of simulating the ensemble of bimodal oscillators. 
 The easiest way to simulate the ensemble on a computer is by updating the state of each oscillator in discrete time steps.

Obtaining the results presented in this paper required a huge computational effort and it was made possible by choosing an efficient way to model the 
system and map the parameter space.  The most direct way to simulate the oscillators is the following: let the model be discretized in time and let $\Delta t$ be the time step.  Then, let the state of each oscillator be stored separately, making the oscillator the basic unit of the model, and in each time step update the state of each oscillator.  This method requires computer time proportional to $N \tmax / \Delta t$ to simulate the dynamics of $N$ oscillators for a time-interval \tmax, i.e.\ the simulation will slow down proportionally to the time resolution $\Delta t$.

A significant speedup can be achieved if the basic unit of the simulation is chosen to be the events that happen to oscillators instead of oscillator states.  Events can be an oscillator turning on (lighting up), an oscillator turning off (darkening), or the start of a new oscillation period.  Events are processed sequentially, in chronological order.  Processing an ``on'' event causes the total output intensity to increase by $1/N$, while an ``off'' event causes it to decrease by $1/N$.  A ``period start'' event causes a new event to be created in the future.  Since there is an upper bound on the time length of a period, events need to be stored only up to a fixed time ahead in the future.  A fixed-length array, containing only the number of each type of event for successive periods of time of length $\Delta t$ can be used for this.  This method requires a time proportional to $\alpha N \tmax + \beta \tmax/\Delta t$, where the constants $\alpha$ and $\beta$ depend on implementation details.  Since in practical implementations $\beta \ll \alpha$, increasing the time resolution does not significantly increase the simulation time, making accurate and fast simulation possible.

\subsection{Mapping the phase space}

Another essential optimization technique used in the present study was sampling the phase space adaptively.  The most common way of mapping the phase space of a system is by simulating the dynamics of the system and calculating the value of the order parameter for each point in a rectangular lattice of points in the phase space.  Increasing the resolution of the lattice twofold causes a $2^n$-fold increase in the number of sample points and consequently a similar increase in the required computation time.

\begin{figure}[bt!]
  \begin{center}
    \includegraphics{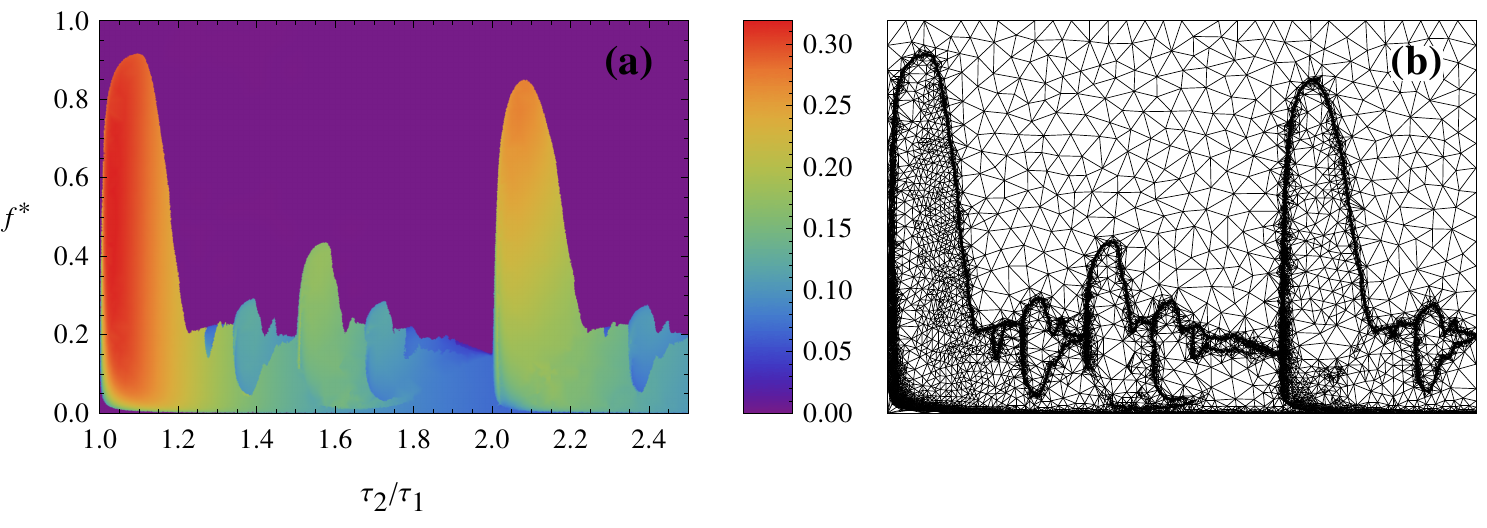}
  \end{center}
  \caption{Phase-space of the model 1 system. The simulation parameters were $N = 10000$ and $\tau_C = 0.15$. (a) The order parameter $\sigma$ as a function of the threshold level $f^*$ and the ratio of the average periods of oscillation modes $\tau_2 / \tau_1$.  The colour indicates the value of $\sigma$: purple regions correspond to no synchronization, for more details see the attached colour-code. (b) Illustration of the adaptively subdivided mesh.}
  \label{fig:mesh}
\end{figure}

A better approach is using adaptive sampling, i.e.\ increasing the number of sample points only in the regions where the behaviour of the order parameter is ``interesting''.  An adaptive sampling method is defined by two choices: 1.\ a quantitative definition of ``interesting'' regions, i.e.\ the criterion for adding more sampling points 2.\ choosing the exact location of new sample points.  Appropriate choices for these should depend on the behaviour of the order parameter as a function of system parameters in the given model. In our system, the parameter space is two-dimensional and the value of the order parameter is constrained to be in the interval $[0,1]$.  The order parameter varies smoothly inside regions that are separated by abrupt and discontinuous transitions, as seen in figure~\ref{fig:mesh}a for the case of model 1.  Due to the Monte Carlo simulation technique used, the results might be noisy.  Based on these considerations, a simple adaptive sampling scheme was chosen.  We start with an arbitrary set of roughly equally spaced sample points and compute the function value (order parameter) in them.  Then in each refinement step, compute the Delaunay triangulation of the point set, and insert a new sample point in the midpoint of each triangle edge if the edge is longer than a threshold and the function values in the two ends differ by more than another threshold.  

This method will trace the shape of the discontinuities very well.  Since most sample points get inserted close to the discontinuities, the increase in the number of points is close to linear than quadratic in the resolution.  Adaptive sampling makes it possible to map the parameter space of the system with high precision with relatively few sample points.  The adaptively subdivided mesh for the case of model 1 is illustrated in figure~\ref{fig:mesh}b. The disadvantage of using such a method is that only those features will be discovered with certainty that have a size comparable to the resolution of the initial mesh.  The features that \emph{are} discovered are mapped with high precision, and the method can produce a detailed looking output, as in figure~\ref{fig:mesh}a.  This may be misleading and one must be aware that the precision of the mapping differs from region to region.

\section{Results}

\begin{figure}[bt]
  \begin{center}
    \includegraphics{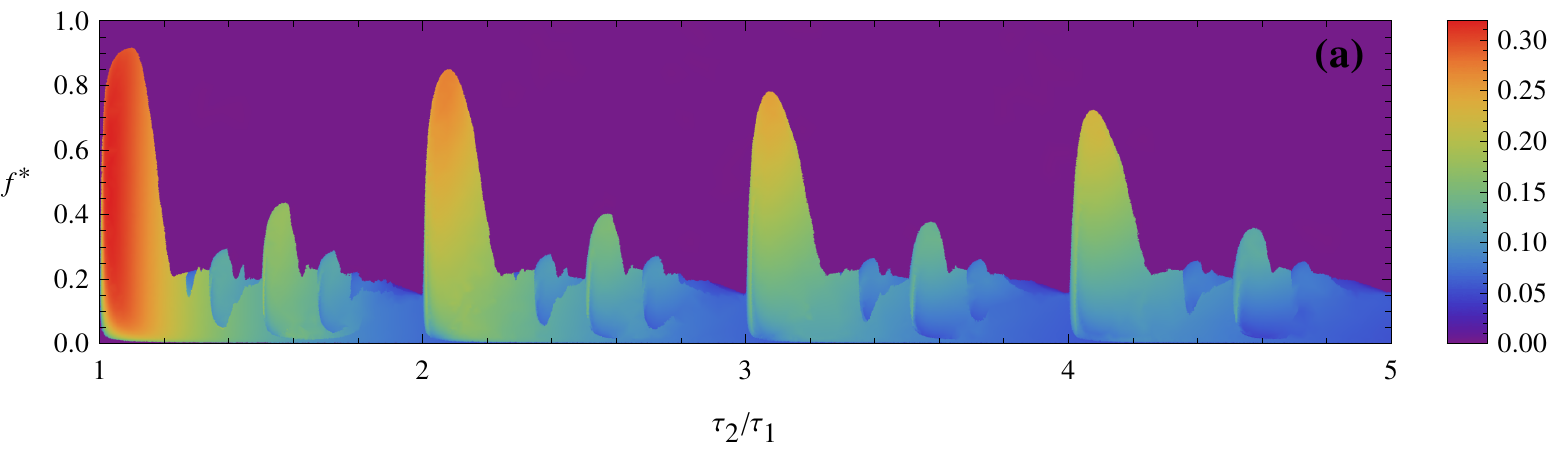}
    \includegraphics{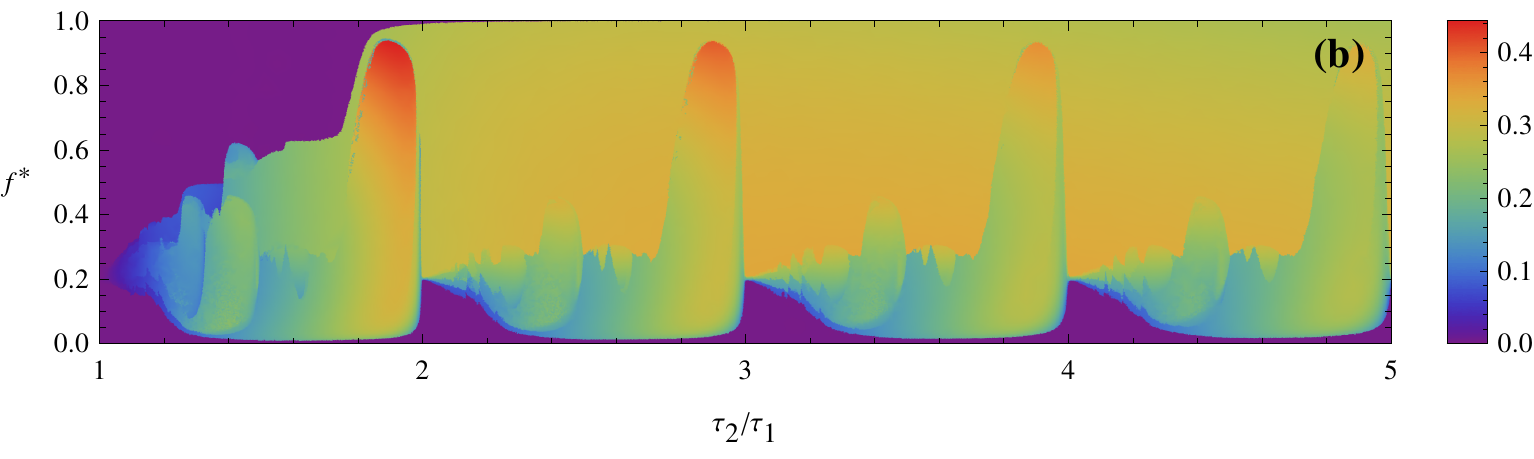}
    \includegraphics{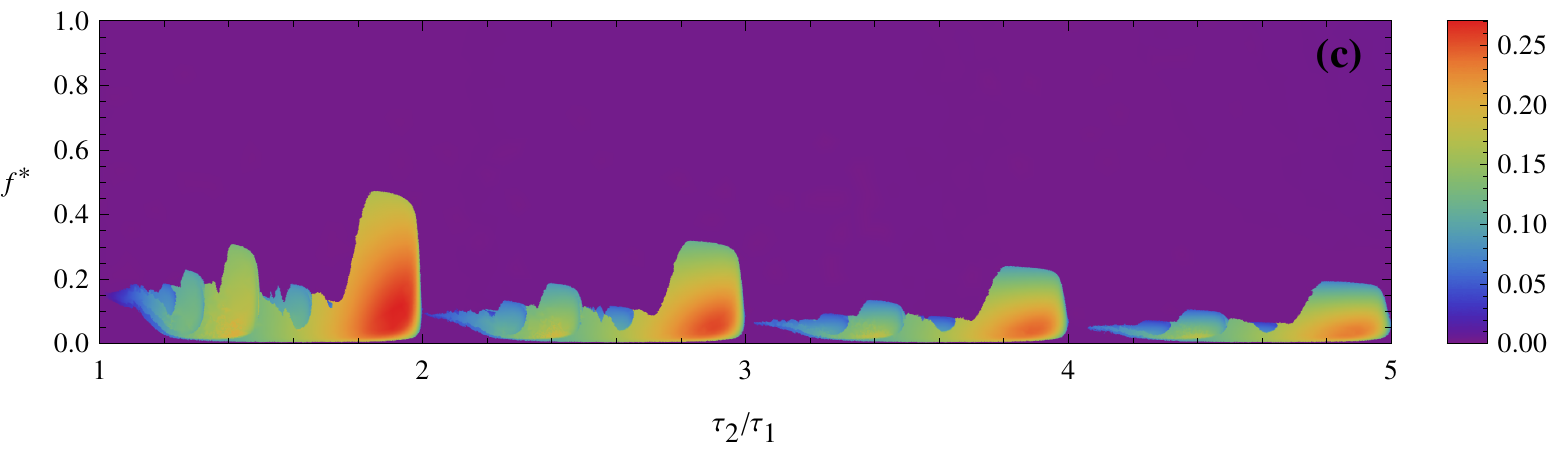}
  \end{center}
  \caption{The phase space of model 1, 2 and 3.  The order parameter $\sigma$ is shown as a function of the threshold level $f^*$ and the ratio of the average periods of oscillation modes $\tau_2 / \tau_1$.  The colour indicates the value of the $\sigma$ order parameter: purple regions correspond to $\sigma = 0$, i.e.\ no synchronization, see also the attached legend.  (a) The phase space of model 1.  The simulation parameters were $N = 10000$ oscillators and $\tau_C = 0.15$.  (b) The phase space of model 2.  $N = 10000$, $\tau_B = 0.8$.  (c) The phase space of model 3.  $N = 10000$, $\tau_C = 0.15$. }
  \label{fig:phase-space-wide}
\end{figure}

Previous numerical studies performed on model 1 and model 2 have found that there is an island-like region of the $f^*$--$\tau^*$ phase-space where synchronization emerges as a side effect of the oscillating units striving to achieve a total output that is close to $f^*$.  Synchronization would occur for the largest interval of $f^*$ when $\tau^*$ (the parameter characterizing the randomness of the periods) was small \cite{Nikitin2001,Neda2003,Sumi2009,Horvat2009}.  The behaviour of the system has not previously been studied as a function of the periods of the oscillation modes, $\tau_1$ and $\tau_2$.  In this paper we focus on mapping the $f^*$ -- $\tau_2/\tau_1$ phase space when $\tau^*$ is much smaller than $\tau_1$ and $\tau_2$.  In order for the system to reach equilibrium, and in order that the stationary state to be independent of the initial state, it is necessary that $\tau^* > 0$.  For all simulations we have fixed the values of $\tau_1$ and $\tau^*$ to be $\tau_1 = 1.0$ and $\tau^* = 0.03$.  It is important to note that this value of $\tau^*$ does not approximate the $\tau^* \rightarrow 0$ limit well.  Reducing its value further will result in some noticeable changes in the structure of the phase space.  However, since the time needed to reach equilibrium was found to grow proportionally with $1/(\tau^*)^2$, the available computation resources imposed a limit on reducing the value of $\tau^*$.  

As described already in section 2, three versions of the model were studied:

\begin{description}
  \item[model 1] is the basic model described in \cite{Nikitin2001,Neda2003}.  The duration of the lit phase, $\tau_C=0.15$, is fixed while the duration of the dark phase, $\tau_B$, can have a greater and a smaller value ($\tau_{B2}$ and $\tau_{B1}$).  The oscillation modes are chosen so as to \emph{optimize} the output intensity $f$ towards $f^*$, i.e.\ minimize the difference between $f$ and $f^*$: if $f \le f^*$ then the short mode ($\tau_1$) is chosen while if $f > f^*$, the long mode is chosen ($\tau_2$).  The mapped phase space is shown in figure~\ref{fig:phase-space-wide}a.

  \item[model 2] has a fixed duration dark phase.  This means that $\tau_B$ is fixed while $\tau_C$ can have a greater and a smaller value, $\tau_{C2}$ and $\tau_{C1}$.  The oscillation modes are chosen to minimize the difference between $f$ and $f^*$ (\emph{optimization}): if $f \le f^*$ then mode 2 (the longer mode) is chosen, increasing the average light intensity; if $f > f^*$ then mode 1 (the shorter mode) is chosen.  The phase space is shown in figure~\ref{fig:phase-space-wide}b. 

  \item[model 3] is similar to model 1, except that the oscillation modes are chosen so as to make $f$ as different from $f^*$ as possible: if $f \le f^*$ then $\tau_2$ is chosen, while if $f > f^*$ then $\tau_1$ is chosen (\emph{anti-optimization}).  The phase space of this model is shown in figure~\ref{fig:phase-space-wide}c.
 
\end{description}

\begin{figure}[hbt]
  \begin{center}
    \includegraphics{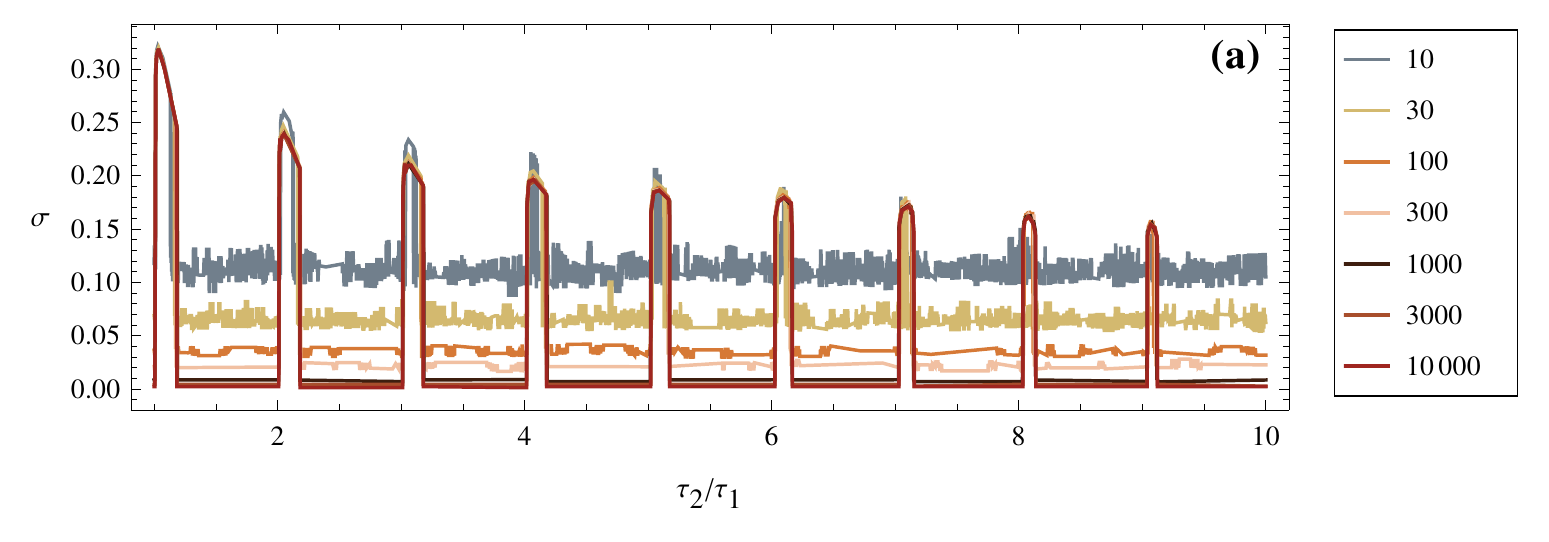}
    \includegraphics{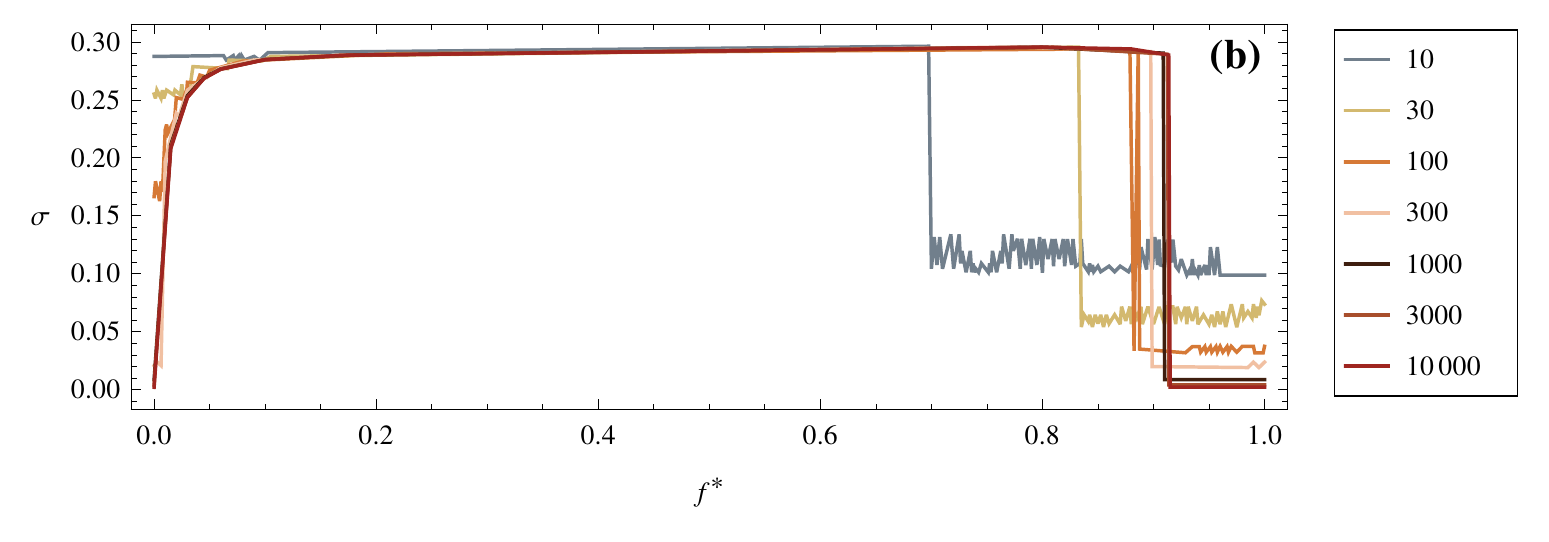}
  \end{center}
  \caption{Finite-size effects for the system. The order parameter $\sigma$ in model 1 as a function of: (a) $\tau_2/\tau_1$ for $f^* = 0.5$; and (b)  $f^*$ for $\tau_2/\tau_1 = 1.1$, for different numbers of oscillators. The curves do not change visibly when the number of oscillators is increased above $N = 3000$. The $\tau_C = 0.15$ value was used.}
  \label{fig:section}
\end{figure}

The time resolution chosen for all simulations presented in the above figures was $\Delta t = 1/1000$.  The simulations were performed up to time $\tmax = 10000$ to ensure that the system reaches a steady state and the order parameter does not change any more.  Then, the period of the signal was estimated using a simple auto-correlation method and the order parameter $\sigma$ was calculated based on an integer number of periods covering approximately the last 500 time units in the data.  The simulations were run for an ensemble containing a number of oscillators ranging from $N=10$ to $N=100{,}000$.  We found that the order-parameter curves do not change significantly when increasing $N$ above 3000. This is nicely visible if one studies finite-size effects for horizontal and vertical sections in figure~\ref{fig:phase-space-wide}a. Characteristic results are shown in figures~\ref{fig:section}a and \ref{fig:section}b, respectively. 

\begin{figure}[p]
  \begin{center}
    \includegraphics[width=0.9\textwidth]{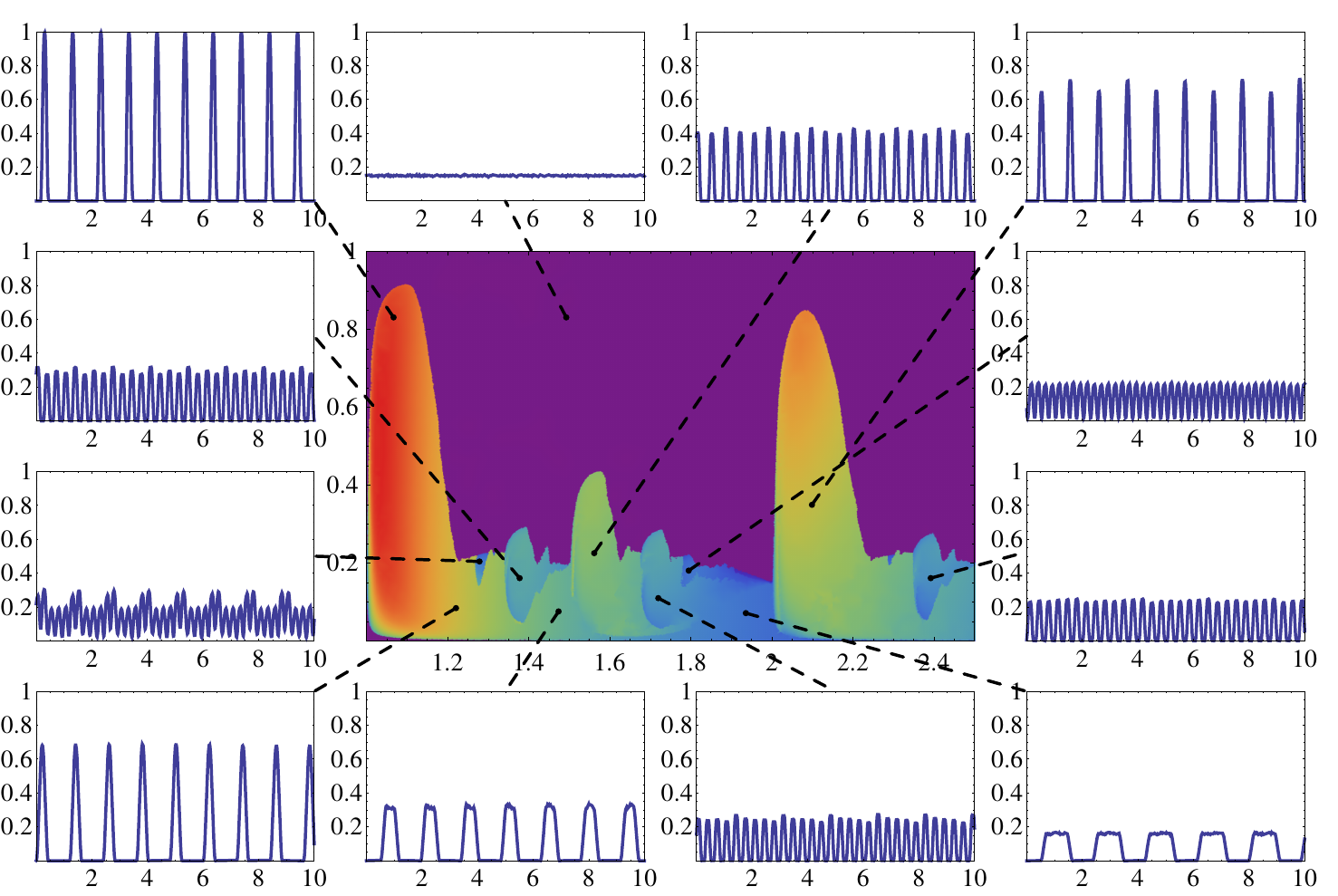}
   \end{center}
  \caption{The shape of the output signal for model 1 is shown for various points in the phase space.  
  ($N = 10000$ oscillators and $\tau_C = 0.15$). The value of the $\sigma$ order parameter is illustrated with the same colour-code as in figure~\ref{fig:phase-space-wide}a. }
  \label{fig:shape1}
\end{figure}

\begin{figure}[p]
  \begin{center}
    \includegraphics[width=0.9\textwidth]{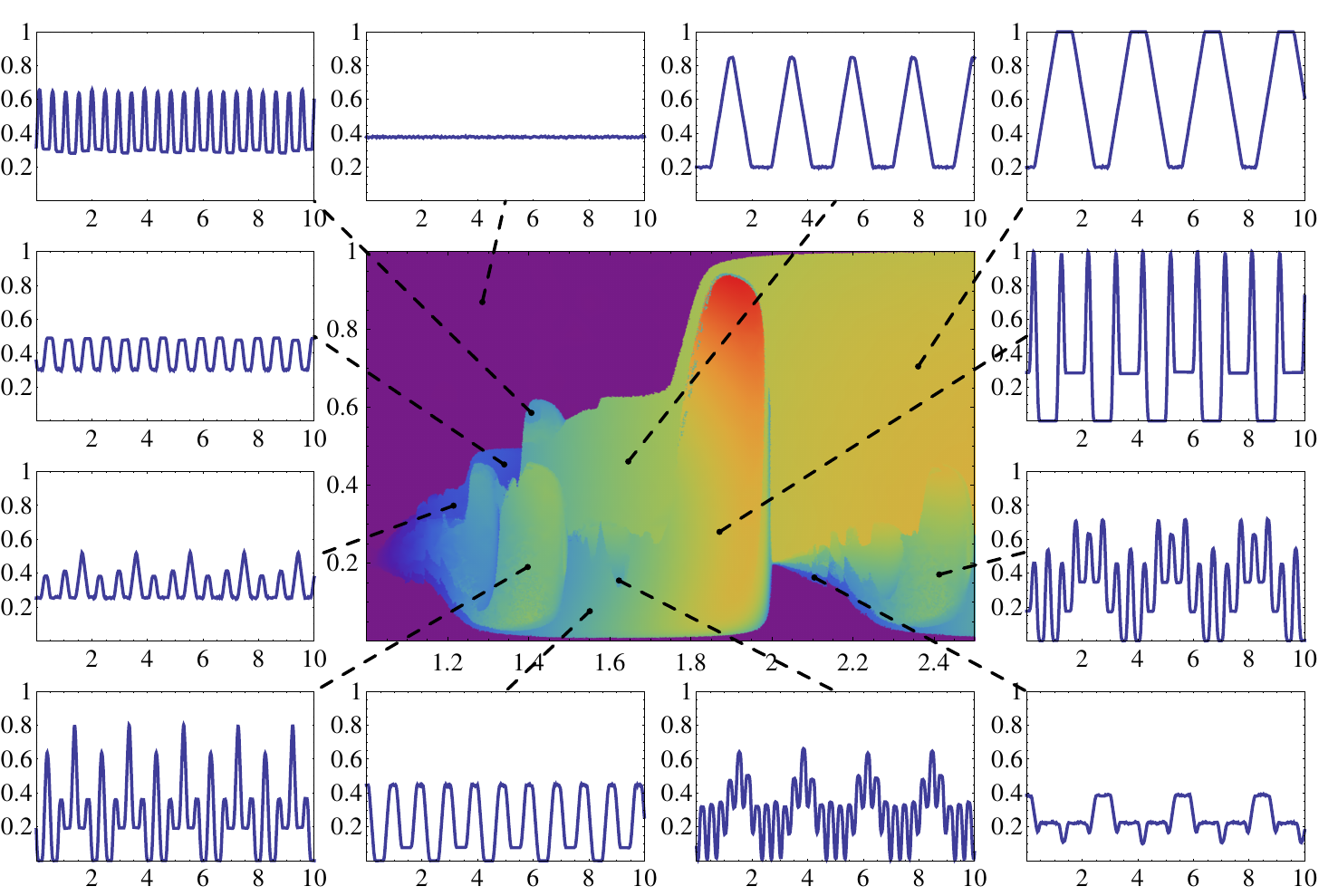}
  \end{center}
  \caption{The shape of the output signal for model 2 is illustrated for various points in the phase space.  
  ($N = 10000$ oscillators,  $\tau_B = 0.8$). The value of the $\sigma$ order parameter is illustrated with the same 
  colour-code as in figure~\ref{fig:phase-space-wide}b. }
  \label{fig:shape2}
\end{figure}

The $f^*$ -- $\tau_2 / \tau_1$ phase space of the system has a complex structure in the case of all three models, and consists of several partially synchronized regions.  The regions are separated by discontinuities in the value of the order parameter $\sigma$.  In each of these regions, the output intensity function of the system, $f(t)$, is periodic but has a different shape.  Some of the shapes of $f(t)$ that occur for different parameter values in model 1 and model 2 are shown in figure~\ref{fig:shape1} and \ref{fig:shape2}. These widely different global signals suggest that the dynamics of this simple system is extremely rich and many different phases are possible. The abrupt appearance of synchronization on the region boundaries and the sudden changes in the shape of the output function resemble phase transitions.

\section{Summary}

Recently, a new type of synchronization model was introduced for pulse-coupled bimodal stochastic oscillators.  This model does not contain an explicit phase difference minimizing force.  Instead, each oscillator chooses a faster or a slower oscillation mode so as to minimize the difference between the system's total output level and a threshold value $f^*$.  Previously this model was studied as a function of the threshold level and the randomness of the oscillation modes.

In this work we have studied three variations of this model, and mapped their behaviour as a function of the threshold level, $f^*$, and the ratio of the average periods of the oscillation modes, $\tau_2 / \tau_1$.  It was found that the ratio of the oscillation modes, which was not considered as a parameter of this model before, has a significant influence on the behaviour of the system.  The $f^*$ -- $\tau_2 / \tau_1$ phase space has a complex structure with several regions, separated by sharp discontinuities of the chosen order parameter.  The shape of the total output intensity function differs between these regions. 

An interesting finding of the present model is that synchronization in such models appears under unexpectedly wide range of conditions. All previous studies have considered a dynamic which intends to minimize the difference between the total output intensity of the system and a threshold level $f^*$.  Here, we have found that synchronization will emerge even when an ``anti-optimizing'' dynamics is used that will maximize the difference between the output level and $f^*$.

The model described in this work is interesting because synchronization emerges not as a result of an explicitly phase difference minimizing interaction, but as a side effect of a simple optimizing (or anti-optimizing) dynamics.  We have found that despite its simplicity, the behaviour of the model changes in an unexpectedly complex manner as a function of the studied parameters, and as a result of this the phase space of the model is also rather complex.

\ack

Work supported from a Romanian IDEI research grant: PN-II-ID-PCE-2011-3-0348.  The work of Sz.~Horv\'at is supported by the European Social Fund through a POSDRU 6/1.5/S/3 2007--2013 PhD grant.  The Triangle fast Delaunay triangulator software was used in the implementation of the adaptive sampling method.

\section*{References}

\bibliographystyle{iopart-num}
\bibliography{sync}

\end{document}